\documentclass{SciPost}
\usepackage{graphicx} % in real document delete demo option
\usepackage{subcaption}
\usepackage{float}

\begin{document}

\begin{center}{\Large \textbf{
Spectral properties of L\'evy Rosenzweig-Porter model via supersymmetric approach
}}\end{center}

% TODO: write the author list here. Use initials + surname format.
% Separate subsequent authors by a comma, omit comma at the end of the list.
% Mark the corresponding author with a superscript *. 
\begin{center}
Elizaveta Safonova\textsuperscript{1,2}, 
Mikhail Feigel' man\textsuperscript{1,3} and
Vladimir Kravtsov\textsuperscript{4}
\end{center}

\begin{center}
{\bf 1} Nanocenter CENN, Ljubljana, Slovenia
\\
{\bf 2} Faculty of Mathematics and Physics, University of Ljubljana, Jadranska 19, 1000 Ljubljana, Slovenia
\\
{\bf 3} Jozef Stefan Institute, Ljubljana, Slovenia
\\
{\bf 4} Abdus Salam International Center for Theoretical Physics,
Strada Costiera 11, 34151 Trieste, Italy
\end{center}

\begin{center}
\today
\end{center}

% For convenience during refereeing: line numbers
%\linenumbers

\section*{Abstract}
{\bf By using the Efetov's super-symmetric formalism we computed analytically the mean spectral density $\rho(E)$ for the L\'evy and the L\'evy -Rosenzweig-Porter random matrices which off-diagonal elements are strongly non-Gaussian with power-law tails. This makes the standard Hubbard-Stratonovich transformation inapplicable to such problems. We used, instead, the functional Hubbard-Stratonovich transformation which allowed to solve the problem analytically for large sizes of matrices.  We show that $\rho(E)$ depends crucially on the control parameter that drives the system through the transition between the ergodic and the fractal phases and it can be used as an order parameter.  
}

\vspace{10pt}
\noindent\rule{\textwidth}{1pt}
\tableofcontents\thispagestyle{fancy}
\noindent\rule{\textwidth}{1pt}
\vspace{10pt}
%%%%%%%%%%%%%%%%%%%%%%%%%%%%%%%%%%%%%%%%%%%%%%%%%%%%%%%%%%%%%%%%%%%%%%%%%%%%%%%%%%%%%%%%%%%%%%%%%%%%%%%%%%%%%%%%%%%%%%%%%%%%%%%%%%%%%
\section{Introduction}
\label{sec:intro}
%%%%%%%%%%%%%%%%%%%%%%%%%%%%%%%%%%%%%%%%%%%%%%%%%%%%%%%%%%%%%%%%%%%%%%%%%%%%%%%%%%%%%%%%%%%%%%%%%%%%%%%%%%%%%%%%%%%%%%%%%%%%%%%%%%%%% 
In recent decade the theory of Anderson localization discovered more than 60 years ago, experienced a certain revival. It is related first of all with the problem of Many-Body Localization (MBL) and absence of thermalization in quantum disordered systems \cite{BAA,Huse_thermalization,Mirlin_MBL}. The second motivation to revisit this problem was experimental realizations   of Hamiltonians  with long-range power-law hopping  in dipolar cold-atom systems   \cite{dipolar1}. By placing such systems in optical cavity one may engineer the power-law hopping with the exponents that are variable in a broad range  \cite{opt-cav}. Both these problems have a common feature:  in the problem of MBL the exponential smallness of the matrix elements connecting two basis states (e.g two bitstrings in a spin chain) at a given Hamming distance (the number of spin flips to get one bitstring from the other) can be compensated by the exponentially large number of such states; in the systems with power-law hopping on a lattice the polynomial smallness of the hopping matrix element between two sites at a large distance can be compensated by the polynomially large number of sites at this distance. The two sites can be simultaneously populated in a given wave functions only if they are in resonance with each other (i.e. the difference in their energies is of the order of or smaller than the transmission matrix element). Therefore, the balance between a small transmission matrix element and a large number of sites  makes significant the probability of such  resonances at large distances between the sites and thus favors sparsely populated but extended wave functions. At a certain range of parameters such wave functions may form a {\it non-ergodic extended state} \cite{DeLuca2014} in which populated sites form a fractal.

In contrast, in the short-range hopping systems on a finite-dimensional lattice, the number of sites grows polynomially but the effective hopping matrix element falls down exponentially at large distance. As the result, in a conventional single-body problem of localization on a lattice only few nearby sites are typically in resonance (localization) at strong disorder or weak hopping; otherwise at large enough hopping or weak disorder all of them are in resonance to form an ergodic state. The intermediate, non-ergodic extended state may realize only at fine tuning of disorder strength, i.e. at the Anderson transition point. In contrast, in interacting systems and in a system on a lattice with power-law hopping such states may form a {\it phase} that exists in  some finite range of parameters.

%%%%%%%%%%%%%%%%%%%%%%%%%%%%%%%%%%%%%%%%%%%%%%%%%%%%%%%%%%%%%%%%%%%%%%%%%%%%%%%%%%%%%%%%%%%%%%%%%%%%%%%%%%%%%%%%%%%%%%%%%%%%%%%%%%%%%%
\begin{figure}[t]
\center{
\includegraphics[width=0.48 \textwidth,angle=0]{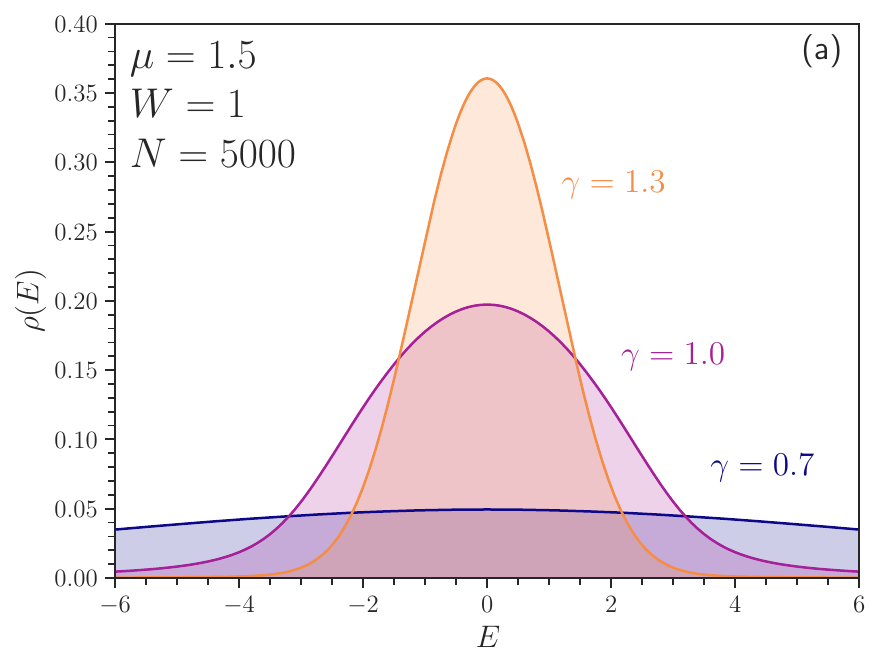}
\includegraphics[width=0.49 \textwidth,angle=0]{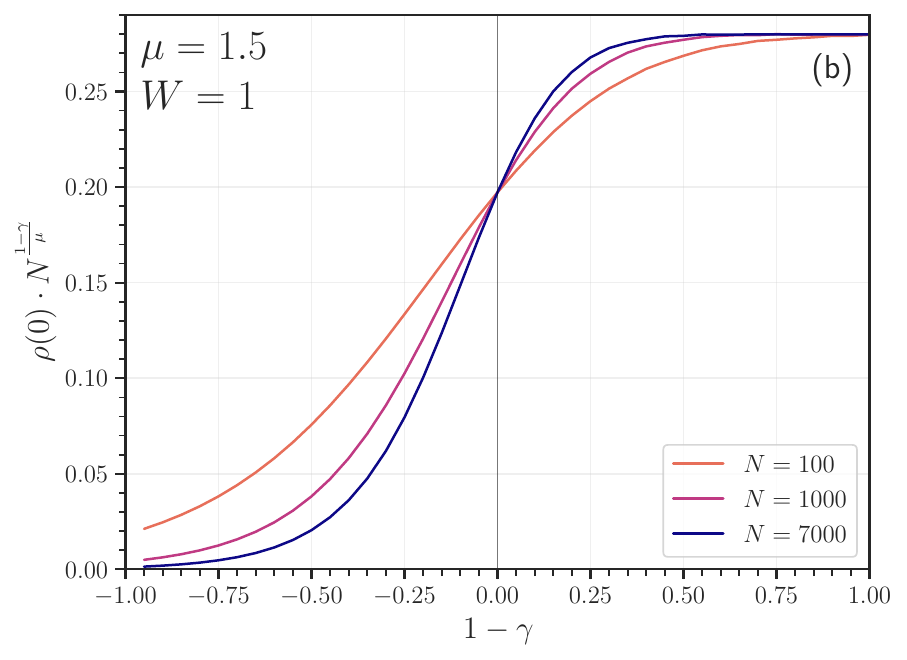}
\includegraphics[width=0.48 \textwidth,angle=0]{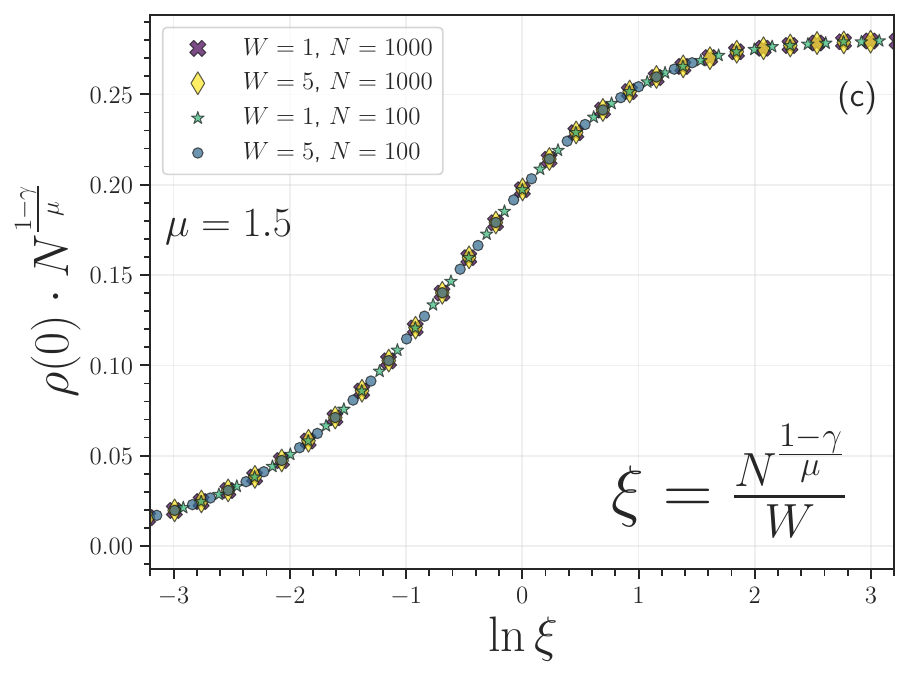}}
\caption{\textbf{DoS $\rho(E)$ at the ergodic transition for the L\'evy-RP random matrices:} (a)  The diverging with $N$  band-width  $B\propto N^{(1-\gamma)/\mu}$ in the ergodic phase, $\gamma=0.7$, the $N$-independent  DoS  at the ergodic transition, $\gamma=1$, and a convergent with increasing $N$ DoS in the non-ergodic extended phase $\gamma=1.3$. (b) Dependence on $N$ of the "order parameter" $\rho(0) N^{\frac{1-\gamma }{\mu}}$ plotted as a function of $\tau=\gamma_{c}-\gamma$. (c) Single-parameter scaling: all curves  for  $\rho(0)N^{\frac{1-\gamma }{\mu}}$ as functions of $\ln \xi = \ln(W^{-1}N^{(1-\gamma)/\mu})$ for different disorder strength $W$ and different matrix sizes $N$ collapse to a single curve which depends only on $\mu$. All plots are obtained  from our analytical results, Eqs.(\ref{f_rescaled}),(\ref{rho_rescaled}).}
\label{fig:main}
\end{figure}
%%%%%%%%%%%%%%%%%%%%%%%%%%%%%%%%%%%%%%%%%%%%%%%%%%%%%%%%%%%%%%%%%%%%%%%%%%%%%%%%%%%%%%%%%%%%%%%%%%%%%%%%%%%%%%%%%%%%%%%%%%%%%%%%%%%%%%
The simple models for such systems are random matrix ensembles with independently fluctuating matrix elements with the distribution functions that depend on the distance from the main diagonal (see various examples in Ref. \cite{Nos}). The simplest of such random matrix ensembles is the   Rosenzweig-Porter (RP) ensemble \cite{RP}, where {\it all} the off-diagonal matrix elements are independently and identically distributed (i.i.d.) according to a certain Gaussian distribution, while the diagonal ones are i.i.d. random Gaussian variables with a different variance. This ensemble is no longer invariant under the basis rotation which makes it a simple playground for studying the localization effects. In particular, one may make the variance of the diagonal matrix elements independent of the matrix size $N$, while the variance of the off-diagonal matrix elements being $N$-dependent  $\propto N^{-\gamma}$ and small ($\gamma>0$). A surprising property of this ensemble  is that it exhibits, besides the localization transition at $\gamma=2$, also a transition at $\gamma=1$ from the ergodic phase identical to the one in the classic Wigner-Dyson random-matrix theory \cite{Mehta}, to the non-ergodic extended phase with the fractal support set of random eigenfunctions \cite{gRP}.  

A further extension of the Rosenzweig-Porter ensemble is the non-Gaussian, strongly tailed distribution of {\it off-diagonal} matrix elements whose higher moments are either fast growing with the matrix size (log-normal RP ensembles \cite{KhayKrav_LN-RP}) or they do not exist   whatsoever (L\'evy-RP \cite{BirTar_Levy-RP}). Such models have a rich phase diagram in the space of two parameters, the {\it hopping parameter} $\gamma$ that determines the typical variance of the hopping matrix elements and {\it the tail parameter} ($p$ \cite{KhayKrav_LN-RP} or $\mu$ \cite{BirTar_Levy-RP}) that controls the fat tails in the distribution of  off-diagonal entries.  

The local dynamics of these models is slow. In particular,   the mean survival probability demonstrates the {\it stretch-exponential} relaxation  of the population of the single initially populated site \cite{LN-RP-RRG21}. This is reminiscent of the corresponding \ stretch-exponential relaxation in the Anderson model on Random Regular Graph \cite{deTomasi2019subdiffusion,KhayKrav_LN-RP,Nosov} and in disordered spin chains \cite{Znidaric2016Diffusive,Scardi16} which is associated with sub-diffusion.

It is important to emphasize that the family of Rosenzweig-Porter random matrix theories is principally different from the theories where {\it all} the entries are independently and identically distributed, for instance from the classic Wigner-Dyson RMT \cite{Mehta} and from the L\'evy matrices \cite{Bouchard_Levy_Mat,Biroli_Levy_Mat}. The presence of the {\it special diagonal} (with the variance of the diagonal matrix elements parametrically larger than that of the off-diagonal ones) breaks the basis invariance and opens the way towards the localization. The difference in the distribution of diagonal and off-diagonal matrix elements makes possible existence of  {\it non-ergodic extended} (fractal) states as a precursor of localization and of the Anderson localization itself. 

The transition from the ergodic to the non-ergodic extended states (the {\it ergodic transition} at $\gamma=\gamma_{ET}$ \cite{gRP,Nos,KhayKrav_LN-RP,BirTar_Levy-RP}) happens as a true phase transition in all the Rosenzweig-Porter models. However, in contrast to the localization transition, it leads to a qualitative change of the {\it global} density of states (DoS) $\rho(E)$. Namely, in the non-ergodic phases the DoS in the thermodynamic limit tends to a distribution of diagonal entries which is $N$-independent, while in the ergodic phase $\rho(E)$ is strongly size-dependent.

We will show in this paper, using an analytic solution for the L\'evy-RP model \cite{BirTar_Levy-RP}, that $\tilde{\rho(0)} = \rho(0) \cdot N^{\frac{1-\gamma }{\mu}}$ may serve as an {\it order parameter} which tends to a non-zero value in the non-ergodic extended phase and tends to zero in the ergodic one. This evolution as a function of $\gamma-\gamma_{ET}=\gamma-1$ depends on the  disorder strength  $W$ (proportional to the square-root of the variance of diagonal matrix elements)  and on the matrix size $N$ combined in a {\it single parameter} 
$\xi=W^{-1}\,N^{(1-\gamma)/\mu}$ which corresponds to the finite-size scaling exponent $\nu=1$ (see Fig.\ref{fig:main}).

To describe  the finite-size scaling (FSS) at the transition from the ergodic to the extended non-ergodic phase one needs the solution for $\rho(E)$ at a finite system size $N$. This solution does not reduce to the solution \cite{math_pure_Levy} for the pure {\it L\'evy ensemble} \cite{Bouchard_Levy_Mat} in which all the matrix elements are identically distributed with the L\'evy distribution. In this work we obtain an exact analytic solution for $\rho(E)$ at a finite (but large) matrix size $N$ using an extension of the Efetov supersymmetry (SUSY) approach \cite{Efetov_book}.

One of the goals of the paper is to demonstrate how the SUSY formalism  can be applied to the random matrices with the heavily tailed, strongly non-Gaussian distribution of off-diagonal entries combined with the different (Gaussian in our case) distribution of the diagonal ones. This method  based on the {\it functional} Hubbard-Stratonovich transformation \cite{MirFyod_non_Gauss, Mirlin:2000cla} can be applied to compute other physical quantities for random Hamiltonians with heavily tailed distribution. The work on  the correlation functions of global and local DoS and the mean survival probability is in progress and the corresponding results will be published elsewhere. \footnote{We would like to note that a popular {\it cavity method} is frequently applied for similar problems leading to various kinds of {\it effective medium approximations}, (see \cite{Bogomol-Giraud} and references therein). However, the accuracy of these methods should be carefully studied for each particular case and the resulting expressions are usually quite involved.}

The rest of the paper is organized as follows. Section \ref{sec:another} provides a detailed description of the model studies and of the functional Hubbard-Stratonovich method which enables us to deal with heavy-tailed distributions using supersymmetric approach. It contains Eqs.(\ref{f(mu, E) integral equation}),(\ref{rho(E) final}) as the main analytical result of the paper. Section~\ref{sec:scaling} is devoted to  scaling analysis of DoS behavior near ergodic transition at $\gamma=1$.  Section \ref{sec:numerics} demonstrates agreement of analytical expressions with results of direct diagonalization. Section \ref{sec:conclu} contains our conclusions.

\section{Density of states for the L\'evy and L\'evy-RP matrices }
\label{sec:another}

\subsection{Definitions}

Our research object is $N \times N$ real symmetric matrix $\hat{H}$ which can be represented as the sum of two symmetric matrices:
\begin{equation}
\hat{H}=\hat{H}^{(1)}+\hat{H}^{(2)},
\end{equation}
where $\hat{H}^{(1)}$ is a diagonal random matrix with independent and identically distributed (i.i.d.) entries and $\hat{H}^{(2)}$ is a full matrix where {\it all} elements are i.i.d. The distribution of $\hat{H}^{(1)}$ and $\hat{H}^{(2)}$ are generally different.
We consider two basic cases:   the case of the {\it L\'evy-Rosenzweig-Porter (L\'evy-RP) matrices} \cite{BirTar_Levy-RP} where the entries of $\hat{H}^{(1)}$ are Gaussian distributed     
\begin{equation}\label{diagonal distribution}
    P_{1}^{(W)}\left(H^{(1)}_{ii}\right)=\frac{1}{\sqrt{2\pi}W}e^{-\frac{H^{(1)2}_{ii}}{2W^2}},
\end{equation}
 and those of $H^{(2)}$ have a   distribution:
\begin{equation}
    P_{2}^{(\mu,\gamma)}\left({H^{(2)}}_{ij}\right)=\left(N^{\gamma}\right)^{1/\mu}L_{\mu}\left(\left(N^{\gamma}\right)^{1/\mu}{H^{(2)}}_{ij}\right),
    \label{P2}
\end{equation}
where $L_\mu (x)$ is a symmetric L\'evy stable distribution  \cite{Levy_fr, Mandelbrot} with the characteristic function:
\begin{equation}\label{char-fun}
    \tilde{L}_{\mu}(k)\equiv \int_{-\infty}^{\infty} L_{\mu}(x) e^{-ikx} dx \equiv e^{-|k|^\mu},\quad 0<\mu\leq2.
\end{equation}
The other case is the {\it L\'evy matrices} \cite{Bouchard_Levy_Mat,Biroli_Levy_Mat}, where {\it all}  entries of  $\hat{H}^{(1)}$ and $\hat{H}^{(2)}$ are independently and identically distributed according to the L\'evy stable distribution   $P_{2}^{(\mu,\gamma)}$ with $\gamma=1$.

Notice that while the variance  $W^{2}$ of $H^{(1)}_{ii}$ is independent of the matrix size $N$, the typical value  of ${H^{(2)}}_{ij}$ scales with $N$ as $N^{-\gamma/\mu}$, and its variance diverges at $\mu<2$ because of the tail   in $L_{\mu}(x)\sim x^{-(1+\mu)}$.
There are two special values of $\mu$: $\mu=2$ where this tail disappears and the distribution $L_{\mu}(x)$ becomes Gaussian, and $\mu=1$ when it coincides with the Cauchy distribution.

Using  Eqs.(\ref{char-fun}),(\ref{P2}) we can find the characteristic function of rescaled $P_{2}^{(\mu,\gamma)}\left({H^{(2)}}_{ij}\right)$ distribution:
\begin{equation}\label{tilde-P_2}
    \tilde{P}_{2}^{(\mu,\gamma)}\left(  k, N\right) = \exp\left(- \frac{\left| k\right|^\mu}{N^\gamma} \right).
\end{equation}

We would like to calculate the mean density of state (DoS) $\rho(E)$ using Efetov's supersymmetric approach \cite{Efetov_book} further elaborated in Refs. \cite{MirFyod_non_Gauss, Mirlin:2000cla}. 

The partition function $Z(E,J)$ in terms of which the mean DoS $\rho(E)$ is found by differentiation over background field $J$, is given in this approach   by the integral over the super-vectors $\phi_{i}$:

\begin{equation}\label{Zfunction}
    Z(E,J) = \int \prod_{i} [d\phi_i] \exp \left\{ \frac{1}{2}\sum_{ij} \phi_i^\dagger \left[ \left( E \hat{I} + J \hat{K}\right) \delta_{ij} - H_{ij}\right] \phi_j \right\},
\end{equation}

\begin{equation}\label{rho}
    \rho\left(E\right)=\frac{1}{2\pi N}\text{Im}\frac{\partial\left\langle Z\left(E,J\right)\right\rangle_{1,2} }{\partial J}\biggr|_{J=0},
\end{equation}

where
\begin{equation}\label{supervector}
    \phi_{i}=\left(\begin{array}{c}
S_{i1}\\
S_{i2}\\
\chi_{i}\\
\chi_{i}^{*}
\end{array}\right),\quad\phi_{i}^{\dagger}=\left(\begin{array}{cccc}
S_{i1} & S_{i2} & \chi_{i}^{*} & -\chi_{i}\end{array}\right)
\end{equation}
is a super-vector with  ordinary (commuting) $\left( S_{i1},\quad
S_{i2} \right)$ and Grassmannian (anti-commuting) $\left(\chi_{i},\quad
\chi_{i}^{*} \right)$ components, $\hat{K}=\text{diag}\left(1,1,-1,-1\right)$ and $\hat{I}$ is identity matrix. We also will need a Grassmannian integration rule:
\begin{equation}\label{Grassmannian integration rule}
    \int \chi d\chi = \int \chi^*d\chi^* =  \frac{i}{\sqrt{2\pi}}.
\end{equation}
%%%%%%%%%%%%%%%%%%%%%%%%%%%%%%%%%%%%%%%%%%%%%%%%%%%%%%%%%%%%%%%%%%%%%%%%%%%%%%%%%%%%%%%%%%%%%%%%%%%%%%%%%%%%%%%%%%%%%%%%%%%%%%%%%%%%%%

%%%%%%%%%%%%%%%%%%%%%%%%%%%%%%%%%%%%%%%%%%%%%%%%%%%%%%%%%%%%%%%%%%%%%%%%%%%%%%%%%%%%%%%%%%%%%%%%%%%%%%%%%%%%%%%%%%%%%%%%%%%%%%%%%%%%%%
%%%%%%%%%%%%%%%%%%%%%%%%%%%%%%%%%%%%%%%%%%%%%%%%%%%%%%%%%%%%%%%%%%%%%%%%%%%%%%%%%%%%%%%%%%%%%%%%%%%%%%%%%%%%%%%%%%%%%%%%%%%%%%%%%%%%%
\subsection{Phase diagram of L\'evy- RP matrices}
\begin{figure}[h!]
\center{
\includegraphics[width=0.48 \textwidth,angle=0]{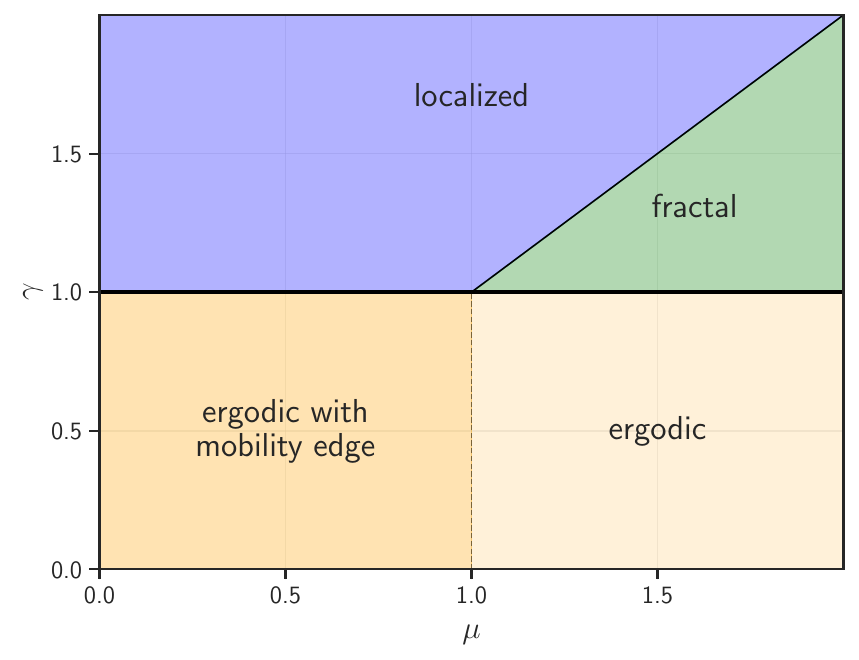}
\caption{\textbf{Phase diagram for the L\'evy-RP model.} }
\label{fig:phase_dia}}
\end{figure}

%%%%%%%%%%%%%%%%%%%%%%%%%%%%%%%%%%%%%%%%%%%%%%%%%%%%%%%%%%%%%%%%%%%%%%%%%%%%%%%%%%%%%%%%%%%%%%%%%%%%%%%%%%%%%%%%%%%%%%%%%%%%%%%%%%%%%%
 Before coming to calculations of the mean DoS, we would like to recall   the main facts about the phase diagram of L\'evy-RP matrices mostly following Ref.\cite{BirTar_Levy-RP}. The phase diagram in the region of interest $0<\mu<2$ is presented in Fig.\ref{fig:phase_dia}. The different phases are identified from the statistics of eigenvectors $\psi_{n}(i)$ which may be ergodic (the inverse participation ratio $I(N)=\sum_{i} |\psi_{n}(i)|^{4}\sim N^{-1}$), localized ($I(N)\sim N^{0}$) or fractal, or extended non-ergodic, ($I(N)\sim N^{-D}$, with $0<D<1$). There are two ergodic (E) phases: the one for $1<\mu<2$ in which eigenvectors are ergodic at all eigenstate energies $E_{n}$ and another one for $0<\mu<1$ where there is a mobility edge (ME) $E_{0}$ beyond which, for  $|E_{n}|>E_{0}$, the eigenvectors are localized.
It is remarkable that all three phases meet at the tricritical point $\gamma=\mu=1$. 

In this paper we are concerned with the simplest spectral statistics which is the mean DoS $\rho(E)$. We show that $\rho(0)$ experience a transition from a $N$-independent value for $\gamma>1$ to the rapidly decreasing with increasing the matrix size $N$ value for $\gamma<1$. According to Fig.\ref{fig:phase_dia} this is associated with the ergodic/non-ergodic transition in the eigenvector statistics. Notice that $\rho(E)$ is insensitive to the transition from the fractal to the localized phase that happens at $\gamma=\mu$, $1<\mu<2$. 

%%%%%%%%%%%%%%%%%%%%%%%%%%%%%%%%%%%%%%%%%%%%%%%%%%%%%%%%%%%%%%%%%%%%%%%%%%%%%%%%%%%%%%%%%%%%%%%%%%%%%%%%%%%%%%%%%%%%%%%%%%%%%%%%%%%%%%
\subsection{Calculation process}
%%%%%%%%%%%%%%%%%%%%%%%%%%%%%%%%%%%%%%%%%%%%%%%%%%%%%%%%%%%%%%%%%%%%%%%%%%%%%%%%%%%%%%%%%%%%%%%%%%%%%%%%%%%%%%%%%%%%%%%%%%%%%%%%%%%%%%

  We start by the averaging of the partition function  $\left( \ref{Zfunction}\right)$ over the random entries of $\hat{H}$:
\begin{equation}\label{<Z>}
\left\langle Z\left(E,J\right)\right\rangle =\int\prod_{i}\left[\phi_{i}\right]
   \exp\left(\frac{i}{2}\sum_{i}\phi_{i}^{\dagger}\left(E+J\hat{K}\right)\phi_{i}+\sum_{i,j}\ln\left\langle \exp\left(-\frac{i}{2}H_{i,j}\phi_{i}^{\dagger}\phi_{j}\right)\right\rangle_{1,2} \right).
\end{equation}
Notation $\langle ...\rangle_{1,2} $ means that averaging is done over both distributions (\ref{diagonal distribution}) and (\ref{P2}).
Since all $H_{ij}$ are not correlated one can split the sum into two terms (diagonal and off-diagonal). Furthermore, because there are $\sim N$ diagonal entries and $\sim N^{2}$ off-diagonal ones,  by setting zero all diagonal entries in the ensemble of L\'evy matrices one obtains only a $1/N$ correction to the pure L\'evy DoS. Neglecting such $1/N$ corrections we can safely obtain the DoS of L\'evy matrices by setting $\gamma=1$ and $W=0$ in the result for the L\'evy-RP ensemble.
\begin{equation}
    \sum_{i,j}\ln\left\langle \exp\left(-\frac{i}{2}H_{i,j}\phi_{i}^{\dagger}\phi_{j}\right)\right\rangle_{1,2} = \sum_{i\neq j}\ln\left\langle \exp\left(-\frac{i}{2}H^{(2)}_{i,j}\phi_{i}^{\dagger}\phi_{j}\right)\right\rangle_{2} + \sum_{i}\ln\left\langle \exp\left(-\frac{i}{2}H^{(1)}_{i,i}\phi_{i}^{\dagger}\phi_{i}\right)\right\rangle_1 .
\end{equation}

 Let us first consider the averaging over the off-diagonal entries of $\hat{H}$. First of all independence of  $H_{i,j}$  allows us to represent the average of products by the product of averages:

\begin{equation}
    \sum_{i\neq j}\ln\left\langle \exp\left(-\frac{i}{2}H^{(2)}_{i,j}\phi_{i}^{\dagger}\phi_{j}\right)\right\rangle_{2} = \ln \prod_{i<j} \left\langle  \exp\left(-i H^{(2)}_{i,j}\phi_{i}^{\dagger}\phi_{j}\right)\right\rangle_{2} = \frac{1}{2}\sum_{i\neq j}\ln\left\langle \exp\left(-i H^{(2)}_ {i,j}\phi_{i}^{\dagger}\phi_{j}\right)\right\rangle_{2}.
\end{equation}

Using smallness $N^{-\gamma/\mu}$   of the typical off-diagonal matrix elements at large enough $N$, we represent:

\begin{equation}\label{ln<exp(-i/2)...> = C/N}
\frac{1}{2}\sum_{i\neq j}\ln\left\langle \exp\left(-\frac{i}{2}H^{(2)}_{i,j}\phi_{i}^{\dagger}\phi_{j}\right)\right\rangle \approx
\sum_{i \neq j}\frac{C\left(\phi_{i}^{\dagger}\phi_{j}\right)}{2N}.
\end{equation}
where we introduced the scalar function
\begin{equation}\label{Cfunction}
    C\left( \phi_i, \phi_j \right) \equiv  C\left( \phi_i^\dagger \phi_j \right) =  N\int dx P_{2}^{(\mu,\gamma)}\left(x,N\right) \left(e^{-i\phi_i^\dagger \phi_j x}-1\right) .
\end{equation}
Thus the averaging over the off-diagonal entries of $\hat{H}$ essentially reduces to the Fourier-transforming of the distribution function  $P_{2}^{(\mu,\gamma)}\left(x,N\right)$.

It is important for further progress of calculations to have a simple enough characteristic function of $P_{2}^{(\mu,\gamma)}$. This was the reason to choose (among all distributions with the power-law tail cut at small values of $x$) the rescaled L\'evy stable distribution for $P_{2}^{(\mu,\gamma)}(x)$. Using Eq.(\ref{tilde-P_2}) for the corresponding characteristic function we obtain for $\gamma>1$ and large enough $N$:
 
\begin{equation}
    C\left( k \right)  = - N^{1-\gamma} \left| k \right|^{\mu}.
\end{equation}

An obvious difficulty that still remains is the non-analytic power $\mu$ of $\phi_{i}^{\dagger}\phi_{j}$ in the functional after averaging over $\hat{H}_{ij}$ (instead of the quartic term arising for the Gaussian distribution). This non-analyticity encodes the fat tails in the distribution which, in their turn, determine the peculiar physical properties of the system. A related problem is that $C\left( \phi_i, \phi_j \right)$  couples the $\phi$- super-vectors at different sites $i$ and $j$. 

In order to decouple the super-vectors we   use {\it the functional Hubbard-Stratonovich(H-S) transformation} instead of the usual one. This non-trivial step was suggested (for different applications)
in Refs.\cite{MirFyod_non_Gauss, Mirlin:2000cla} and rarely used since then \footnote{see, however, two recent works \cite{Akara-Evnin, Evnin} where this formalism was successfully applied to the problem of certain sparse matrices}. Since this mathematical trick is a common framework for treating all the random Hamiltonians with a fat tail in the distribution, we present it here in detail. 

The conventional Hubbard-Stratonovich (HS) transformation converts the quartic term in the action into the quadratic (or bi-linear) one which can be easily integrated, even if the fields for different sites are coupled in the bi-linear form. When the distribution of off-diagonal matrix elements are strongly non-Gaussian, an essentially non-linear (and non-quartic!) expression of coupled fields emerges after the averaging over the random potential. Such an expression cannot be reduced to the quadratic one by any integral transformation. Instead, the fields corresponding to different sites can be first made decoupled by the functional HS transformation. The resulting local term in the action, albeit non-linear, can be treated within the saddle-point approximation. This saddle-point approximation is exact up to $1/N$ corrections which are known to exist also in the Gaussian case on the top of the semi-circle spectral density. At the same time, the finite-size effects proportional to $N^{(1-\gamma)/\mu}\gg 1/N$   are taken into account. Those effects are especially important in the vicinity of the ergodic transition when $(1-\gamma)/\mu$   are small.

We start from the identity:

\[\exp\left(\frac{1}{2N}\int\left[d\psi\right]\left[d\psi'\right]v\left(\psi\right)C\left(\psi,\psi'\right)v\left(\psi'\right)\right)= \]

\begin{equation}\label{<z> before saddle point}
    \int Dg\exp\left(-\frac{N}{2}\int\left[d\psi\right]\left[d\psi'\right]g\left(\psi\right)C^{-1}\left(\psi,\psi'\right)g\left(\psi'\right) +\int\left[d\psi\right]g\left(\psi\right)v\left(\psi\right)\right),
\end{equation}
where $C\left(\psi,\psi'\right)$, $v(\psi)$ and $g(\psi)$ are some functions or fields.

We suppose that $C^{-1}$ operator exists and determined by the relation

\begin{equation}
\int\left[d\chi\right]C^{-1}\left(\phi,\chi\right)C\left(\chi,\psi\right)=\delta_{\phi,\psi},
\end{equation}
where $\delta_{\phi,\psi}$ is $\delta-$function in the space of supervectors. Choosing $v\left(\psi\right)=\sum\limits_{i=1}^{N}\delta\left(\psi-\phi_{i}\right)$ we will get an expression in which $\phi_{i}^{\dagger}$ and $\phi_{j}$ are decoupled for different $i,j$:

\begin{equation}\label{after decoupling}
   \exp\left(\sum_{i,j}\frac{C\left(\phi_{i}^{\dagger}\phi_{j}\right)}{2N}\right)=\int Dg \exp\left(-\frac{N}{2}\int\left[d\psi\right]\left[d\psi'\right]g\left(\psi\right)C^{-1}\left(\psi,\psi'\right)g\left(\psi'\right)+\sum_{i}g\left(\phi_{i}\right)\right).
\end{equation}

After substitution Eq.(\ref{after decoupling}) into Eqs.(\ref{<Z>})-(\ref{ln<exp(-i/2)...> = C/N})  and averaging over $\hat{H}^{(1)}$, the partition function takes the form:

\[\left\langle Z\left(E,J\right)\right\rangle =\int Dg\exp\left(-\frac{N}{2}\int\left[d\psi\right]\left[d\psi'\right]g\left(\psi\right)C^{-1}\left(\psi,\psi'\right)g\left(\psi'\right) \right. \]

\begin{equation}
    \left. +N\ln\left\{ \int\left[d\phi\right]\exp\left(\frac{i}{2}\phi^{\dagger}\left(E+J\hat{K}\right)\phi+g\left(\phi\right) +\ln\left(\int P_{1}^{(W)}\left(\epsilon\right)e^{-\frac{i}{2}\epsilon\phi^{\dagger}\phi}d\epsilon\right)\right)\right\} \right).
\end{equation}
Here we suppress the site indices in $\phi$ as after decoupling  the integration over all $\phi_{i}$ are independent and identical, thus resulting in only the pre-factor $N$ in front of the result of integration over one of them denoted by $\phi$.

Performing the functional integration over $g$  by the steepest descent method (justified by a large pre-factor $N$ in the action), we get the following integral equation for $g$(for $J=0$):

\begin{equation}\label{integral equation g(phi)}
   g\left(\psi\right)= \frac{\int\left[d\phi\right]C\left(\psi,\phi\right)\exp\left(\frac{i}{2}\phi^{\dagger}E\phi+g\left(\phi\right)+\ln\left(\int P_{1}^{(W)}\left(\epsilon\right)e^{-\frac{i}{2}\epsilon\phi^{\dagger}\phi}d\epsilon\right)\right)}{\int\left[d\phi\right]\exp\left(\frac{i}{2}\phi^{\dagger}E\phi+g\left(\phi\right)+\ln\left(\int P_{1}^{(W)}\left(\epsilon\right)e^{-\frac{i}{2}\epsilon\phi^{\dagger}\phi}d\epsilon\right)\right)}.
\end{equation}
This equation does not change under the unitary super-vector rotation $\phi \rightarrow \hat{T}\phi$, $\psi \rightarrow \hat{T}\psi$  since the r.h.s. of (\ref{integral equation g(phi)}) contains only $\phi^\dagger\phi$ and $\psi^{\dagger}\phi$ combinations. For this reason the solution for $g(\phi)$ is invariant under this rotation.  We therefore search for a solution $g\left(\phi \right)$ as a function of the invariant $\phi^\dagger \phi \equiv \mathbf{S}^2 + 2\chi^* \chi$. 

The integral in the denominator of Eq.(\ref{integral equation g(phi)}) has an integrand in which the super-symmetry is not violated.  Thus due to the basic property of the super-symmetry method \cite{Efetov_book} it is equal to 1.  
Then introducing the components, Eq.(\ref{supervector}), of a super-vector explicitly, we express $g\left(\phi\right)$ in a form 
\begin{equation}
    g\left( \phi \right) \equiv g_0 \left( \phi^\dagger \phi\right) = g_0 \left( \mathbf{S}^2\right) + 2\chi^* \chi g_0' \left( \mathbf{S}^2\right).
\end{equation}

For the future calculation let us introduce a scalar function of commuting variables:
\begin{equation}
    F\left(R^{2}\right)\equiv\frac{i}{2}ER^{2}+g_0\left(R^{2}\right)+\ln\left(\int P_{1}^{(W)}\left(\epsilon\right)e^{-\frac{i}{2}\epsilon R^2 }d\epsilon\right).
\end{equation}
If we separate the commuting part from both sides of the integral equation Eq.(\ref{integral equation g(phi)}) we obtain
\begin{equation}
    g_0\left(\mathbf{S}^2\right) = -N^{1-\gamma} \int \frac{2dR_1 dR_2}{-2\pi} \left| \mathbf{SR} \right|^\mu e^{F\left(\mathbf{R}^{2}\right)} \frac{\partial F \left(R^2\right)}{\partial R^2}.
\end{equation}
where $R^2 = R_1^2 + R_2^2$.

Here we used the definition, Eq.(\ref{supervector}),  and the Grassmann integration rule (\ref{Grassmannian integration rule}). After switching to the polar coordinates and integration by parts this expression takes the form:

\begin{equation}
    g_0\left( x \right) = -x^{\mu/2}\frac{2 N^{1-\gamma}}{\mu B\left( \frac{1}{2},\frac{\mu}{2}\right)}\int_0^\infty dy \exp\left( F\left( y^{2/\mu}\right) \right),
\end{equation}
where $B\left( \frac{1}{2},\frac{\mu}{2}\right)$ is the $\beta$-function.

If we search the   solution $g_0\left( x\right)$ in the form $g_0(x) \equiv - f_{\mu}(E) x^{\mu/2}$ we  arrive at  an integral equation for $f_{\mu}(E)$.

\begin{equation}\label{f(mu, E) integral equation}
    f_{\mu}(E) = \frac{ N^{1-\gamma}}{ B\left( \frac{1}{2},\frac{\mu}{2}\right)}\int\limits_{0}^{\infty} \frac{dy}{y^{1-\mu/2}} \exp\left( i\frac{E}{2} y - f_{\mu}(E) y^{\mu/2} + \ln\left( \int P_{1}^{(W)}\left(\epsilon\right)e^{-\frac{i}{2}y\epsilon}d\epsilon \right) \right).
\end{equation}
Notice that this equation is not a true integral equation but rather an ordinary (transcendental) equation, as the  function $f_{\mu}(E)$ is not integrated.

Now it is time to return to the DoS calculation. Since we know saddle-point solution of (\ref{integral equation g(phi)}) we can write down the expression for $\rho(E)\propto {\rm Im}\frac{\partial Z}{\partial J}\biggr|_{J=0}$ in the large-$N$ limit. Using (\ref{<z> before saddle point}) we obtain:

\begin{equation}
     \frac{\partial\left\langle Z\left(E,J\right)\right\rangle }{\partial J}\biggr|_{J=0}=\frac{iN}{2}\int\left[d\phi\right]\exp\left(\frac{i}{2}E\phi^{\dagger}\phi+g\left(\phi\right)+\ln\left(\int P_{1}^{(W)}\left(\epsilon\right)e^{-\frac{i}{2}\phi^\dagger \phi \epsilon}d\epsilon\right)\right)\phi^{\dagger}\hat{K}\phi.
\end{equation}

Performing Grassmannian integration and using integration by parts we  arrive at the final result
\begin{equation}\label{rho(E) final}
    \rho\left(E\right)=\text{Re}\left[\frac{1}{2\pi}\int_{0}^{\infty}dye^{\frac{i}{2}Ey-f_{\mu}\left(\frac{E}{2}\right)y^{\mu/2}+\ln\left(\int P_{1}^{(W)}\left(\epsilon\right)e^{-\frac{i}{2}y\epsilon}d\epsilon\right)}\right],
\end{equation}
where $f_{\mu}(E)$ should be extracted from Eq.(\ref{f(mu, E) integral equation}).

Eqs.(\ref{f(mu, E) integral equation}),(\ref{rho(E) final}) is the main result of this paper. It is valid in the limit of large (but finite) $N$ both for a pure L\'evy matrices \cite{Bouchard_Levy_Mat,Biroli_Levy_Mat} (corresponding to $W=0, \gamma=1$) and for the L\'evy-RP matrices with the special diagonal.  In particular,  it works for the L\'evy-RP matrices \cite{BirTar_Levy-RP} with the Gaussian weight of diagonals, where
\begin{equation}
    \ln\left(\int P_{1}^{(W)}\left(x\right)e^{-\frac{i}{2}y x}dx\right) =  -\frac{W^2 y^2}{8}.
\end{equation}
The result for the pure L\'evy matrices  is known in the mathematical literature \cite{math_pure_Levy}. In this case $\rho(E)$ is $N$-independent in the large-$N$ limit. It was obtained using the traditional mathematical tools and the proof of the formula equivalent to Eqs.(\ref{f(mu, E) integral equation}),(\ref{rho(E) final})  involves rather long chain of arguments.  The power of  supersymmetric calculus used in the present paper makes it possible to reach the same result much faster.

Notice also that the result of Ref.\cite{Bouchard_Levy_Mat} for $\rho(E)$ for L\'evy matrices is not identical to our result (e.g. it requires  a solution for two unknown functions) though numerically they are very close.

The mean DoS for the more interesting case of L\'evy-RP matrices \cite{BirTar_Levy-RP} where there are both the localization and the ergodic transitions and a non-trivial fractal phase, is a totally new result. It allows to study a non-trivial and $N$-dependent variation of $\rho(E)$ as $\gamma$ crosses the ergodic transition at $\gamma=1$. 

Last but not least, the derivation in the framework of the Efetov's supersymmetric approach presented above contains elements common to all problems of  random Hamiltonians with a heavy tailed distribution of parameters and thus it is quite general.
%%%%%%%%%%%%%%%%%%%%%%%%%%%%%%%%%%%%%%%%%%%%%%%%%%%%%%%%%%%%%%%%%%%%%%%%%%%%%%%%%%%%%%%%%%%%%%%%%%%%%%%%%%%%%%%%%%%%%%%%%%%%%%%%%%%
\section{Single-parameter scaling}
\label{sec:scaling}
%%%%%%%%%%%%%%%%%%%%%%%%%%%%%%%%%%%%%%%%%%%%%%%%%%%%%%%%%%%%%%%%%%%%%%%%%%%%%%%%%%%%%%%%%%%%%%%%%%%%%%%%%%%%%%%%%%%%%%%%%%%%%%%%%%%
The DoS for the L\'evy-RP matrices depends both on the strength $W$ of the diagonal disorder and on the typical value of the hopping (off-diagonal) matrix elements controlled by the parameter $\gamma$. However, at large enough matrix size $N$ this dependence is in fact a dependence of a single-parameter:
\begin{equation}\label{xi}
\xi=W^{-1}\,N^{\frac{1-\gamma}{\mu}}.
\end{equation}

 To show that  one can make a rescaling  
\begin{eqnarray}
E &\rightarrow& \epsilon\, N^{\frac{1-\gamma}{\mu}},\\ \nonumber
f_{\mu}(E) &\rightarrow& Y_{\mu}(\epsilon)\, N^{\frac{1-\gamma}{2}},\\ \nonumber
y &\rightarrow& t\, N^{-\frac{1-\gamma}{\mu}}.
\end{eqnarray}
Then Eqs.(\ref{f(mu, E) integral equation}),(\ref{rho(E) final}) take the form:

\begin{equation}\label{f_rescaled}
     Y_{\mu}(\epsilon) = \frac{ 1}{ B\left( \frac{1}{2},\frac{\mu}{2}\right)}\int\limits_{0}^{\infty} \frac{d t}{t^{1-\mu/2}} \exp\left( \frac{i}{2}\,\epsilon\, t -  Y_{\mu}(\epsilon)\, t^{\mu/2} - \xi^{-2}\,\frac{t^2}{8} \right),
\end{equation}
\begin{equation}\label{rho_rescaled}
     \tilde{\rho}\left(\epsilon\right)=\text{Re}\left[\frac{1}{2\pi}\int_{0}^{\infty}dt\,\exp\left(\frac{i}{2}\,\epsilon\, t-
		Y_{\mu}\left(\epsilon\right) t^{\mu/2}- \xi^{-2}\, \frac{t^2}{8}\right)\right], \quad \tilde{\rho}\left(\epsilon\right) = \rho(E) \cdot N^{\frac{1-\gamma}{\mu}}.
\end{equation}

Now it is clear that: 
\begin{itemize}

\item the rescaled DoS $\tilde{\rho}(\epsilon) $ obeys a single-parameter scaling, i.e. it depends on a single parameter $\xi$ rather than on $W$ and $N$ separately.
\footnote{For the Gaussian Rosenzweig-Porter ensemble this result was obtained recently by the replica method \cite{Cugliandolo}.}
\item For L\'evy matrices (corresponding to $W=0, \gamma=1)$ $\rho(E)$ is an $N$-independent function of the energy $E$.
\item For L\'evy- RP matrices ($W\neq 0$) and $\gamma<1$ (the ergodic phase) the DoS  $\rho(E)=N^{-\frac{(1-\gamma)}{\mu}}\rho_{\text{L\'evy}}\left(E\,N^{-\frac{(1-\gamma)}{\mu}}\right)$ 

converges in $N\rightarrow\infty$ limit to that for the L\'evy matrices with the rescaled energy,  while for $\gamma>1$ (fractal or localized phase) it converges to the Gaussian distribution of the diagonal matrix elements. At the ergodic transition $\gamma=1$ $\rho(E)$ is $N$-independent but depends on the diagonal disorder $W$ and $\mu$.
\end{itemize}
The inflation of the body of the distribution (and the corresponding decrease of $\rho(0)$ by normalization) in the ergodic phase of L\'evy-RP model is illustrated in Fig.\ref{fig:main}(a). The single-parameter scaling is reflected in the perfect collapse of data for different $N$ and $\gamma$ in Fig.\ref{fig:main}(c).

The single-parameter scaling allows to find the critical exponent $\nu$ of finite-size scaling (FSS) at the ergodic transition $\gamma=1$. By definition of FSS any quantity, e.g. $\tilde{\rho}(0)$, near the $\gamma$-driven transition  must obey at $|\gamma-1|\ll 1$  the scaling relation:
\begin{equation}
\tilde{\rho}(0)= R_{\mu}\left(L^{\frac{1}{\nu}}\,(\gamma-1)\right),
\end{equation}
where $R_{\mu}(x)$ is a scaling function that depends only on $\mu$, and $L$ is a properly defined length scale. For all Rosenzweig-Porter matrices $L=\ln N$. In a particular case of L\'evy-RP matrices the single parameter scaling suggests that the dependence is only on $\xi={\rm exp}[\ln N \,(1-\gamma)/\mu]$, which implies that the function $R_{\mu}$ depends on the combination $L\,(1-\gamma)$, where $L=\ln N$, and on $\mu$. From that it immediately follows that the exponent $\nu$ at the ergodic transition is equal to:
\begin{equation}
\nu=1.
\end{equation} 

\section{Limiting cases and numerical  verification}
\label{sec:numerics}

\subsection{Pure L\'evy ensemble   }
 
In this section we verify our analytical result, Eqs.(\ref{f(mu, E) integral equation}),(\ref{rho(E) final}), by exact numerical diagonalization and averaging over the ensemble of corresponding random matrices. We start by the case of L\'evy matrices that corresponds to 
\texorpdfstring{$\gamma = 1, W=0$}{gamma = 1, W=0}.

A general receipt is to solve Eq.(\ref{f(mu, E) integral equation}) numerically, then substitute it in Eq.(\ref{rho(E) final}) and compare the result with the result of numerical diagonalization of random matrices. However in some trivial cases Eqs.(\ref{f(mu, E) integral equation}), (\ref{rho(E) final}) allow for an analytical solution. In particular, the case $\mu=2$ represents the textbook example of the Gaussian Orthogonal Ensemble (GOE) \cite{Mehta}.  In this case Eq.(\ref{f(mu, E) integral equation}) reduces to the quadratic equation with the solution for $\rho(E)$ in a form of celebrated semi-circle 
\footnote{L\'evy stable distribution at $\mu=2$ reproduces the normal distribution with variance $\sigma=\sqrt{\frac{2}{N}}$ This is why the eigenvalues are normalized differently and one can expect the spectra from $-2\sqrt{2}$ to $2\sqrt{2}$ instead of $(-2,2)$ in case of normal distribution with $\sigma=\sqrt{\frac{1}{N}}$.}
\begin{equation}
    f_{2}(E) = \frac{1}{2}\left( i\frac{E}{2} \pm \sqrt{2-\frac{E^2}{4}} \right)\Rightarrow \rho_{\text{GOE}}(E)\biggr|_{\mu=2} = \frac{1}{2\pi}\sqrt{2-\frac{E^2}{4}}.
\end{equation}

\begin{figure}[t]
    \centering
    \includegraphics[width=0.32 \textwidth]{ 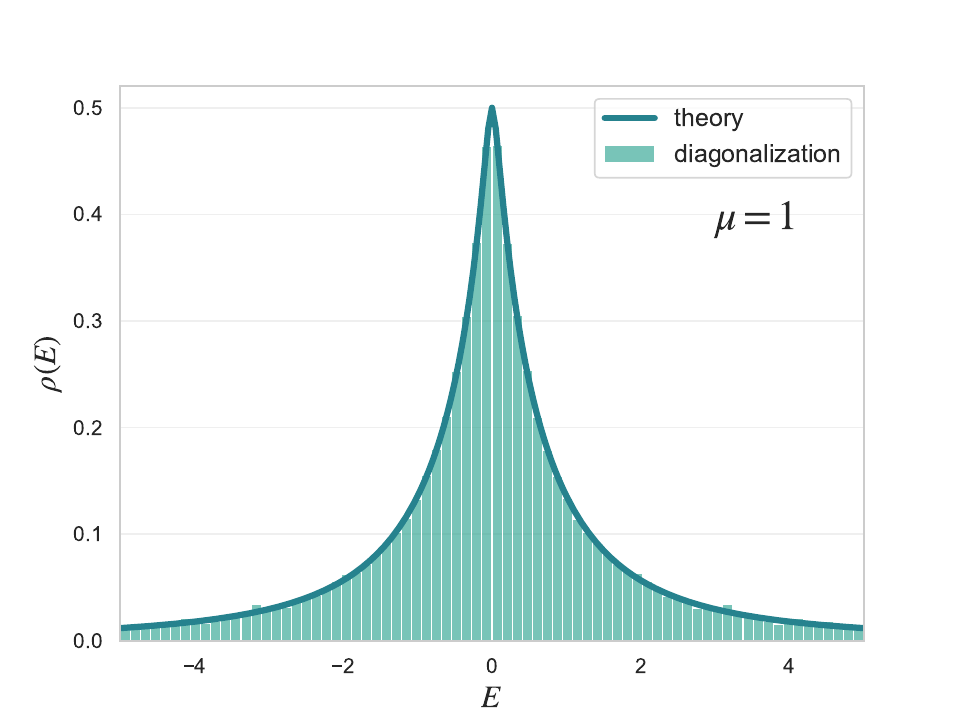}
    \includegraphics[width=0.32 \textwidth]{ 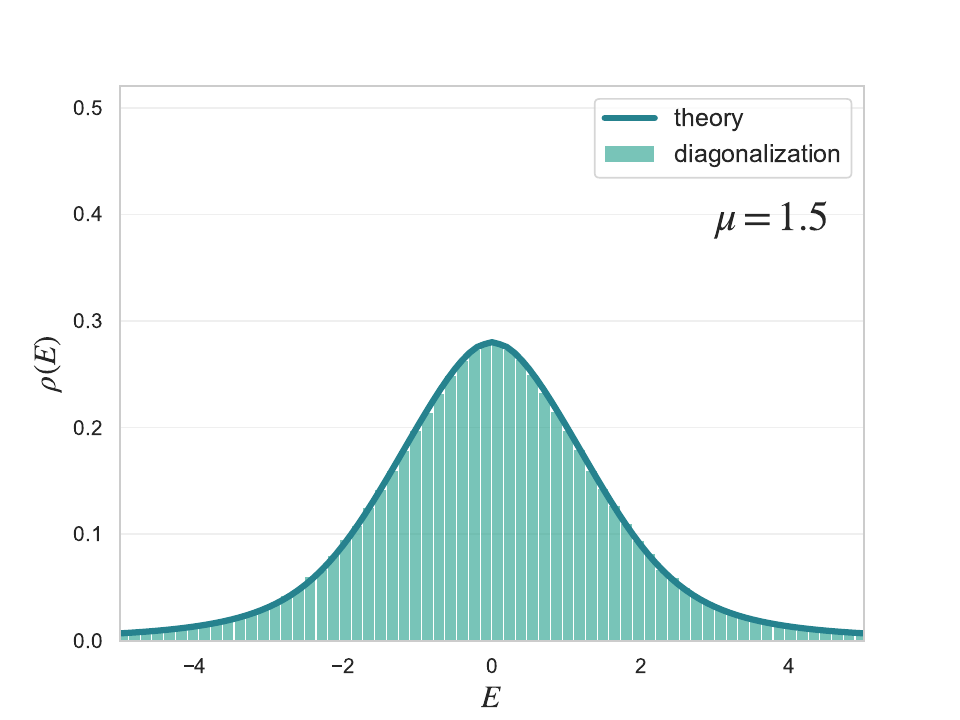}
    \includegraphics[width=0.32 \textwidth]{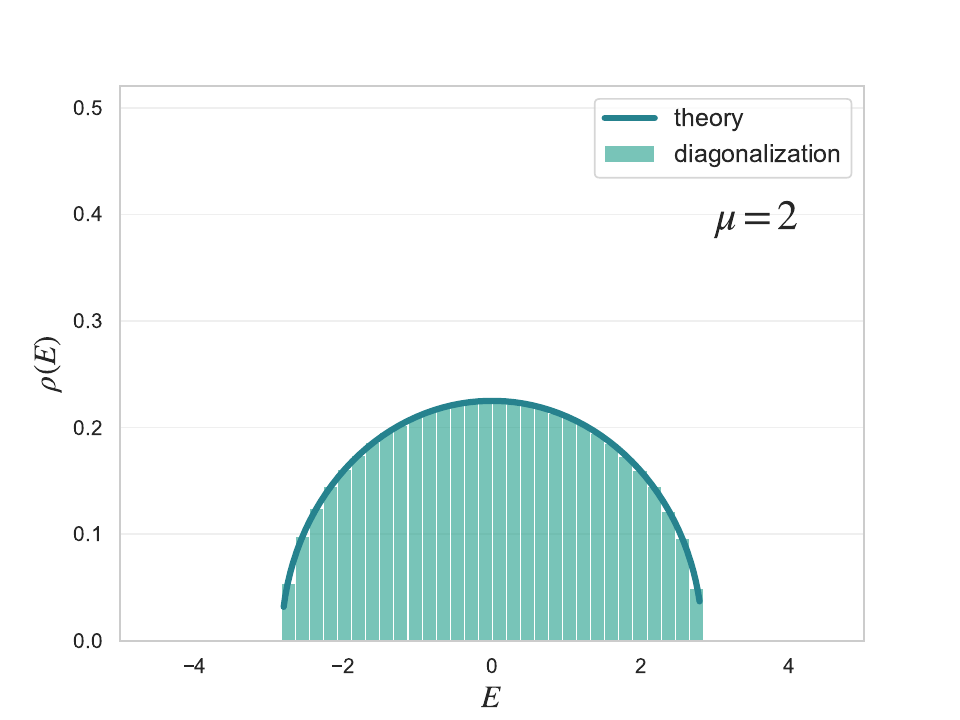 }
    \caption{Comparison of the analytical results of Eqs.(\ref{f(mu, E) integral equation}),(\ref{rho(E) final}) (shown by green lines)   with the numerical diagonalization of the L\'evy matrices for $N=5000$(10 different samples) and $\mu=1$(Cauchy), $\mu=3/2$, $\mu=2$(Gauss) (shown by histograms).}
	\label{fig:Levy}
\end{figure}

For an arbitrary $0<\mu<2$ some analytical results are also possible. In particular, it could be useful to calculate $\rho_{\text{L\'evy}}(E=0)$:
\begin{equation}\label{rho0 Levy}
    f_{\text{Levy}}(\mu,0) = \sqrt{ \frac{2}{\mu B\left( \frac{1}{2},\frac{\mu}{2} \right)} } \Rightarrow \rho_{\text{Levy}}(0) = \frac{\Gamma\left( \frac{2}{\mu} +1 \right)}{2\pi} \left[ \frac{\mu  B\left( \frac{1}{2},\frac{\mu}{2} \right) }{2} \right]^{1/\mu}.
\end{equation}
However, the goal of this section is to compare the analytical results with the results of numerical diagonalization, in order to establish how well the saddle-point approximation, Eq.(\ref{integral equation g(phi)}), works at a reasonably large matrix sizes $N\sim 5000$. One can see the accuracy of this approximation in Fig.\ref{fig:Levy}.
%%%%%%%%%%%%%%%%%%%%%%%%%%%%%%%%%%%%%%%%%%%%%%%%%%%%%%%%%%%%%%%%%%%%%%%%%%%%%%%%%%%%%%%%%%%%%%%%%%%%%%%%%%%%%%%%%%%%%%%%%%%%%%%%%%%%%
%%%%%%%%%%%%%%%%%%%%%%%%%%%%%%%%%%%%%%%%%%%%%%%%%%%%%%%%%%%%%%%%%%%%%%%%%%%%%%%%%%%%%%%%%%%%%%%%%%%%%%%%%%%%%%%%%%%%%%%%%%%%%%%%%%%
 \begin{figure}[t]
    \centering
    \includegraphics[width=0.32 \linewidth]{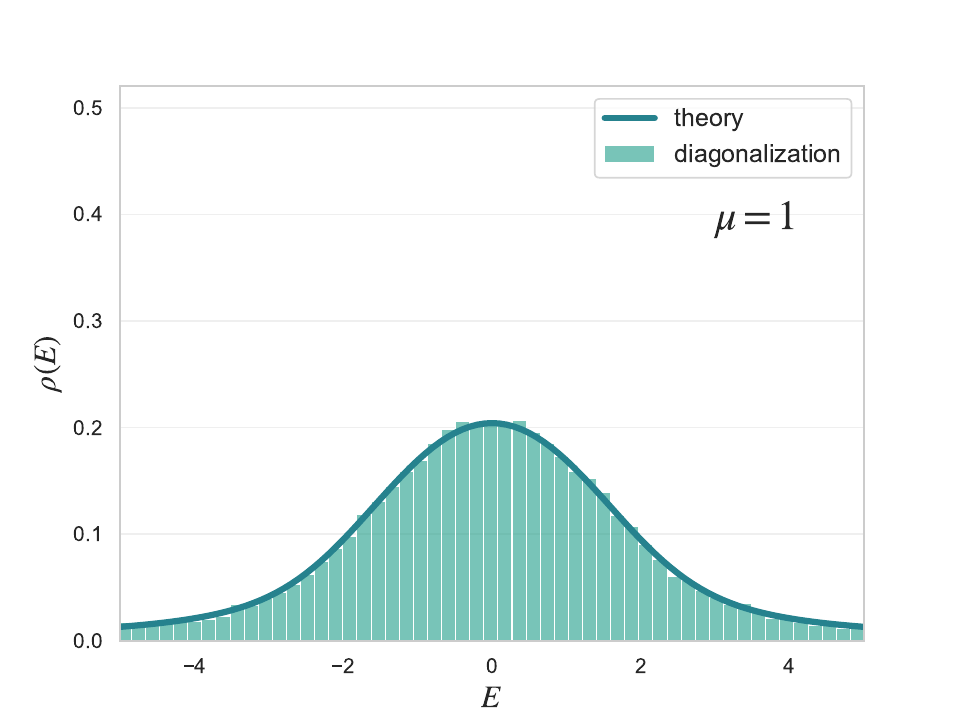 }
    \includegraphics[width=0.32 \textwidth]{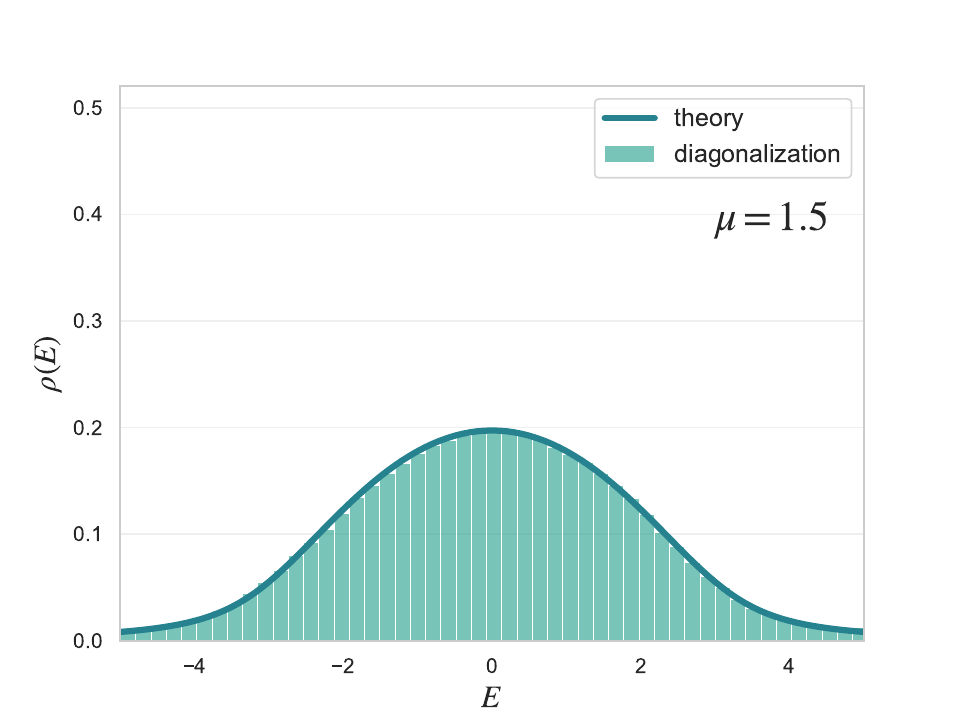}
		\includegraphics[width=0.32 \textwidth]{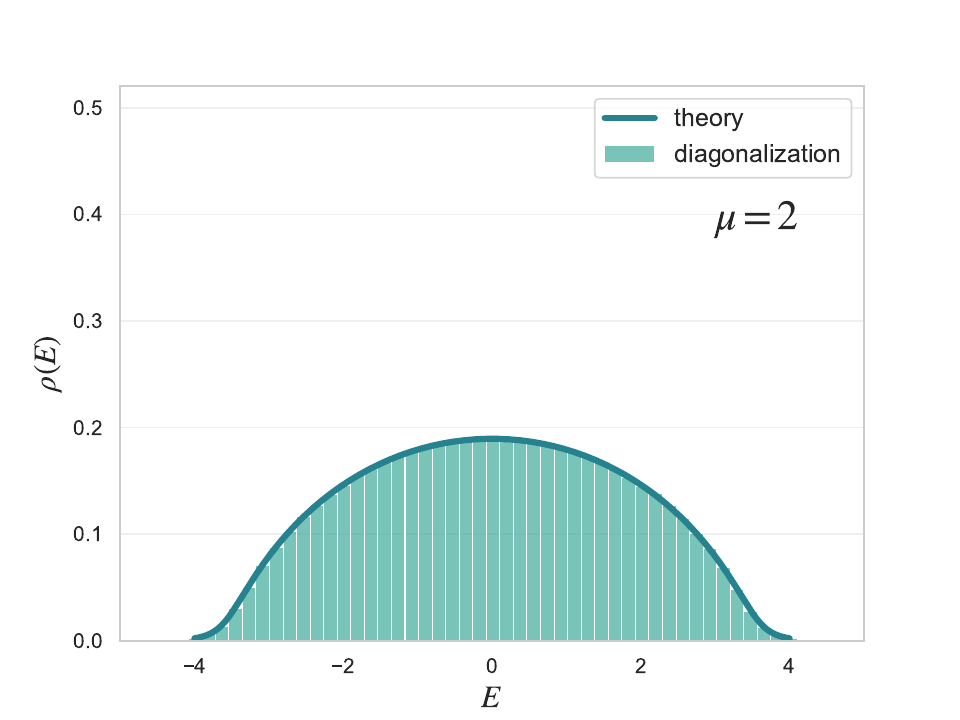}
    \caption{Comparison of the analytical results of Eqs.(\ref{f(mu, E) integral equation}),(\ref{rho(E) final}) (shown by green lines)   with the numerical diagonalization of the L\'evy-RP matrices for $\gamma=1$, $W=1$, $N=5000$(10 samples) and $\mu=1$,  $\mu=1.5$,  $\mu=2$ (shown by histograms).  }
	\label{fig:Levy-RP}
\end{figure}
%%%%%%%%%%%%%%%%%%%%%%%%%%%%%%%%%%%%%%%%%%%%%%%%%%%%%%%%%%%%%%%%%%%%%%%%%%%%%%%%%%%%%%%%%%%%%%%%%%%%%%%%%%%%%%%%%%%%%%%%%%%%%%%%%%%%
 
\subsection{L\'evy-RP ensemble}
Now we  perform a similar comparison for the L\'evy-RP matrices at $\gamma=1$ and $W=1$. The choice of $\gamma=1$ is the most non-trivial case, as for $\gamma>1$ the $N\rightarrow \infty$ limit of $\rho(E)$ coincides with the Gaussian distribution of the diagonal elements, while for $\gamma<1$ it coincides with the rescaled $\rho(E)$ for pure L\'evy matrices. The results for the three  values of $\mu$ ($\mu=2$, $\mu=1$ and $\mu=3/2$) are shown in Fig.\ref{fig:Levy-RP}. As for pure L\'evy matrices, the comparison demonstrates an excellent coincidence of the analytical results and those of numerical diagonalization. Also clear is the hybrid character of the distribution: for $\mu=2$ the band edge is no longer sharp as for a semi-circle, with appearance of the Gaussian tails; for $\mu=1$ the cusp at $E=0$ is rounded as for the Gaussian distribution.
%%%%%%%%%%%%%%%%%%%%%%%%%%%%%%%%%%%%%%%%%%%%%%%%%%%%%%%%%%%%%%%%%%%%%%%%%%%%%%%%%%%%%%%%%%%%%%%%%%%%%%%%%%%%%%%%%%%%%%%%%%%%%%%%%%%%% 
\section{Conclusion}
\label{sec:conclu}
%%%%%%%%%%%%%%%%%%%%%%%%%%%%%%%%%%%%%%%%%%%%%%%%%%%%%%%%%%%%%%%%%%%%%%%%%%%%%%%%%%%%%%%%%%%%%%%%%%%%%%%%%%%%%%%%%%%%%%%%%%%%%%%%%%%%%
In this work we show how to obtain spectral statistics of random matrices with heavily tailed distribution of elements within the Efetov's super-symmetry formalism. We consider two important examples: the pure L\'evy symmetric matrices where all the elements are i.i.d. with the L\'evy $\mu$-stable distribution and the L\'evy-Rosenzweig-Porter matrices, where the diagonal elements are i.i.d. with the  Gaussian distribution  and the off-diagonal elements are i.i.d. with the L\'evy $\mu$-stable distribution and a small typical value that scales with the matrix size $N$ as $\sim N^{-\gamma/\mu}$. The fact that the diagonal matrix elements are $\sim N^{0}$ and the off-diagonal elements are typically small for large matrices results in a rich phase diagram with the ergodic, fractal and the localized phases and transitions between them. By computing the mean spectral density we show that it is sensitive to the transition between the ergodic and the fractal non-ergodic   states (the ergodic transition), with the properly rescaled maximal spectral density $\tilde{\rho}(0)$ behaving like an order parameter for such a transition. Furthermore, we have shown that the dependence of $\tilde{\rho}(0)$ on the matrix size and the strength of  disorder reduces to the dependence on a single parameter $\xi$.  From the dependence of this parameter on the matrix size $N$ and the control parameter $\gamma$ that drives through the ergodic transition we found that the critical exponent of the finite-size scaling for this transition is $\nu=1$.

All the analytically obtained results are verified by exact numerical diagonalization of matrices of large sizes. 

The mathematical formalism employed in computing analytically the mean spectral density within the Efetov's super-symmetry approach is quite general and can be applied to different problems of random Hamiltonians which parameters have broad distributions with heavy tails. 

\section*{Acknowledgments}  
We are grateful to Yan Fyodorov and Aleksey Lunkin for numerous useful discussions.

\bibliography{Levy-RP}
%%%%%%%%%%%%%%%%%%%%%%%%%%%%%%%%%%%%%%%%%%%%%%%%%%%%%%%%%5

\nolinenumbers

\end{document}